# A Review on Cyber Crimes on the Internet of Things

Mohan Krishna Kagita*, Navod Thilakarathne, Thippa Reddy Gadekallu, Praveen Kumar Reddy Maddikunta, and Saurabh Singh

[1] School of Computing and Mathematics,Charles Sturt University, Melbourne, Australia.
[2] Department of ICT, University of Colombo,Sri Lanka.
[3] School of Information Technology and Engineering, Vellore Institute of Technology, India.
[4] Department of Industrial and System Engineering, Dongguk university, Seoul, South Korea.
mohankrishna4k@gmail.com, navod.neranjan@ict.cmb.ac.lk, thippareddy.g@vit.ac.in, praveenkumarreddy@vit.ac.in, saurabh89@dongguk.edu

**Abstract.** Internet of Things (IoT) devices are rapidly becoming universal. The success of IoT can't be ignored in today's scenario; along with its attacks and threats on IoT devices and facilities are also increasing day by day. Cyber-attacks become a part of IoT and affecting the life and society of users, so steps must be taken to defend cyber seriously. Cybercrimes threaten the infrastructure of governments and businesses globally and can damage the users in innumerable ways. With the global cybercrime damages predicted to cost up to 6 trillion dollars annually on the global economy by cyber-crime. Estimated of 328 Million Dollar annual losses with the cyber-attacks in Australia itself. Various steps are taken to slow down these attacks but unfortunately not able to achieve success properly. Therefor secure IoT is the need of this time and understanding of attacks and threats in IoT structure should be studied. The reasons for cyber-attacks can be 1. Countries having week cyber securities,2. Cybercriminals use new technologies to attack. 3. Cybercrime is possible with services and other business schemes. MSP (Managed Service Providers) face different difficulties in fighting with Cyber-crime. They have to ensure that customer's security as well as their security in terms of their servers, devices, and systems. Hence, they must use effective, fast, and easily usable antivirus and anti-malware tools.

---

*

# 1   Introduction

IoT (Internet of things) is developing very rapidly and it offers various types of services that made it the fastest-growing technology with a big influence on society and business infrastructures. IoT has become an integral part of human's modern life like in education, every type of business, healthcare, stores the sensitive data about companies and individuals, information about financial transactions, development of the product, and its marketing[37] [49]. In IoT, Transmission from connected devices has generated huge demand to concentrate on security as millions and billions of users perform sensitive transactions on the internet. Cyber threats and attacks are rising daily in both complexity and numbers. Potential attackers are increasing with the growth in networks and also the tools or methods they are using are becoming more effective, efficient, and sophisticated[15]. Hence, to get the full potential of IoT, it is needed to be protected from threats and attacks. Smart devices or technologies like a hot spot, internet other IoT has entered in every part of the life of human beings, and security is compromised[20].

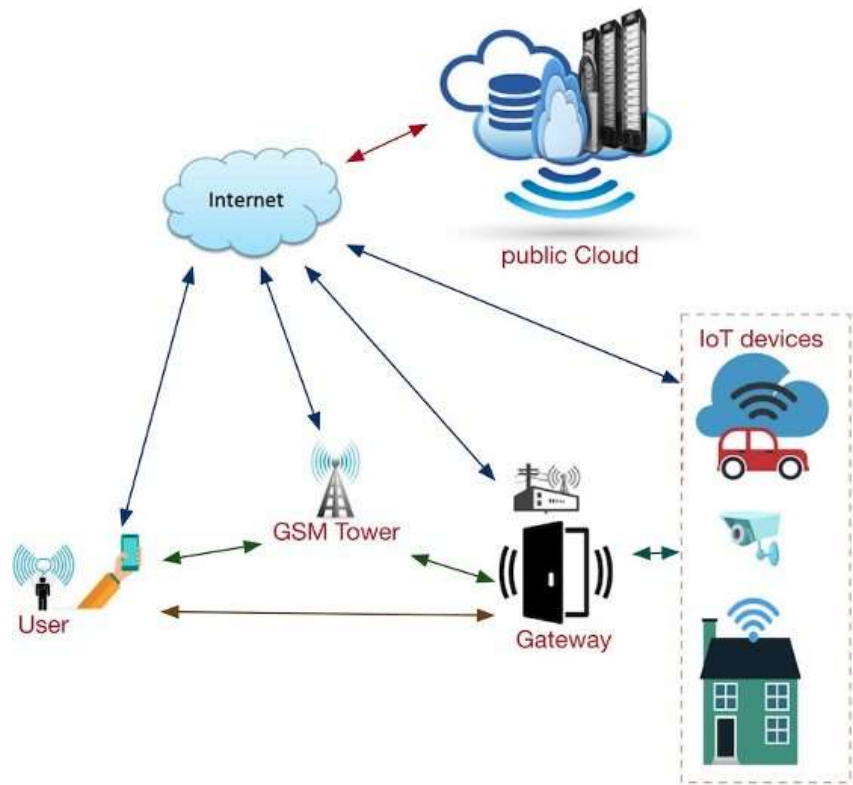

**Fig. 1.** A high level of System Model of IoT

These smart technologies indeed have many advantages to offer but still, it has many loopholes that result in the possibility of cyber-attacks that will cause huge damage to life and property. In today's scenario, it becomes essential that technologies with proper security systems should be adopted in the whole ecosystem to finish the possibilities of hacking or fraud. 98% of all the IoT devices traffic is un-encrypted, exposing personal and confidential data on the network[27]. The most used IoT device in the bossiness and every day in the office space is IP-Phones which has 44% of enterprise IoT Devices but has only 5% of security issues when compared to other IoT devices. Most security issues are faced by cameras that have 33% of risk but only 5% of usage in the business world[21][22].

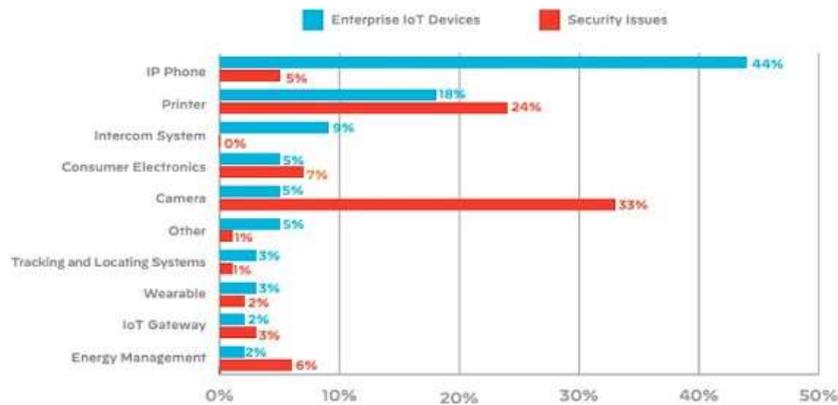

**Fig. 2.** Industrial usage of IoT devices

As per the reports of Australia cybersecurity center, there has been one reported incident for every 10 minutes, which costs Australian economies $328 million annual losses. The Top five Cybercrime types targeting Australians are Identity theft, online fraud, and shopping scams, Bulk extortion, Online romance scams, Wire-fraud, and business email compromise[26]. The period of December 2019 to June 2020 Agari data finds 68% of the identity-deception based attacks aimed at impersonating a trusted individual or a brand. As per the reports from Kaspersky estimate of 105 Million attacks on IoT devices are coming from 276,000 unique IP addresses. Cybercriminals uses network to infect smart devices to conduct DDos attacks as a proxy services. 51% of health care devices attacked

involves imaging devices[53]. New techniques , such like peer-to-peer command and worm-like features for self-propagation, are used to infect vulnerable IoT devices on same network[50].

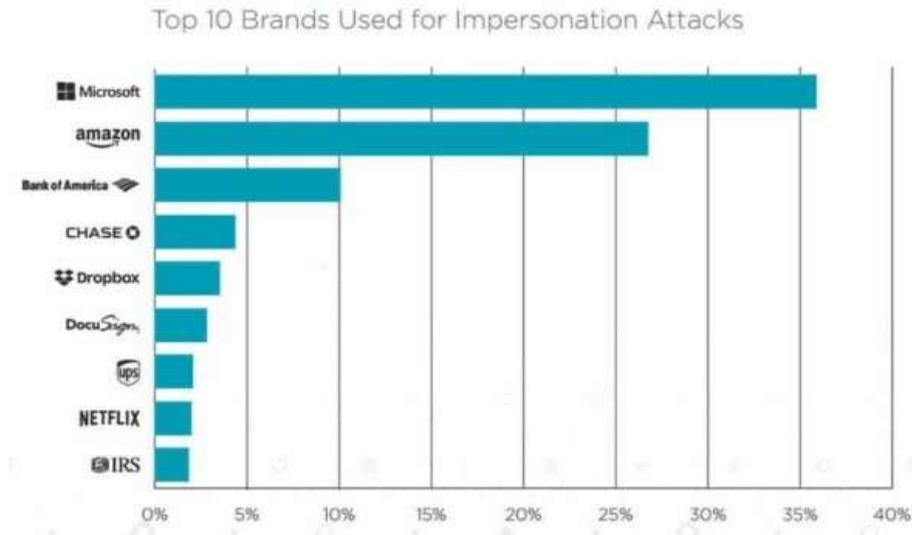

**Fig. 3.** Email Fraud and Identity Deception Trends by Agari

## 2    LITERATURE REVIEW

**Yang Lu and Li Da Xu,** (2019) explore that IoT (Internet of things)  modernized the network globally including smart equipment, Humans, intelligent objects, and information. IoT provides an opportunity to increase integrity, confidentiality, accessibility, and availability. Internets of things are still on the developing stage and contain many issues that are necessary to be solved[1]. System security is a base for expansion in IoT. In this study, cybersecurity is studied thoroughly. The main focus is on the protection and integration of various smart devices and technologies in information communication. This study helps other researchers and experts who want to do researches in IoT in the future can get useful information[39]. The study shows the research of cybersecurity on IoT, taxonomy, and architecture of IoT cybersecurity, strategies, and other trends in research and challenges.

**Hilt, S.; Kropotov, V.; Merces, F.; Rosario, M.  and  Sancho,  D.** (2017) deliberate that IoT is influencing every field of today's society. The ongoing progress of IoT becomes the temptation for cybercriminals[2]. Various studies are done on how criminals can attack IoT and what impact will it laid on. In this research IoT cyber-crime underground is selected to collect ideas about the current fears from different minds that consider them. 5 underground groups are analyzed which are categorized based on the language used in discussions among community i.e. English, Spanish, Portuguese, Russian and Spanish. Many research and tutorials are gathered on the method of hacking, exploitation of vulnerability but no symbol of any intensive determination from the criminal's clusters to enormous damage of IoT Structures are found. Cybercriminals are usually inspired by financial profit and until now there are few ways of making money from the attack on IoT. New devices are invented by cybercriminals with which they find new ways to infect and make a lot of money with these infections.

**Dr. Venckauskas, A.; Dr. Damasevicius, R.; Dr. Jusas, V.; Dr. Toldinas, J.; Rudzika, D. and Dregvaite, G**. (2015), found that IoT can be described as a physical system of world wide web that unites all types of physical things on the internet. The Internet of things is quite complex[3]. The big size, scope, and vast physical distribution of IoT make it difficult to protect it from threats and attacks of cybercrimes[38]. Limitations of the Internet of Things like low power add a contribution to the problems by prohibited the usage of high security but techniques of resource greedy cryptography[23]. IoT provides a stage for cyber-crimes. Hackers or attackers take benefit of a low level of understanding among users of IoT technologies and safety measures to cheat them. New methods for digital forensics in IoT should be increased as the number of threats and attacks of existing and futuristic will increase day by day. New technologies and new strategies are continuously developing that create new challenges for digital forensics. IoT has a large amount of information. Huge scope, a large number of data, diverse nature of the Internet of things, methods in which information is shared, combines and handled need the inventions of new strategies by the digital forensic examinations. It is found that traditional forensic methods used in investigating cyber-crimes that are not effective at all, as new tools and devices are developed by cybercriminals to defraud the user on IoT[32].

**Abomhara, M. and Koien, G. M.** (2015) found that in the field of IoT security, still lots of work to be done by merchants and end-users. It is necessary to find out the imitations of IoT security as the internet on things is growing so fast. Threats and attacks of IoT infrastructure should be thoroughly studied along with the consequences of these threats and attacks against IoT[4]. It is advised that effective security devices for controlling access, authentication, manage identity, flexible faith management structure should be taken care of at the starting of the development of the product. This research is useful for the researchers who study the field of security by helping in finding the mains problems insecurity of IoT and provide a clear understanding of threats and attacks and the factors instigating from various organizations or intelligence agencies. This research helps in better clarity on various threats and their factors inventing from various burglars such as intelligence and organizations[28]. The procedure of finding threats and vulnerabilities is important to make healthy and com-pletely secured IoT and make sure that the security solution is secure enough to protect from malicious attacks.

**J. Iqbal and B. M. Beigh,**( 2017), explore that as the use of the internet is increasing in the world, cyber-crime is also increasing at the same pace, particularly in India[5]. IoT covers vast areas all over the world in the same way Cyber-crime is not bound to a specific geographical area. So, these crimes can't be controlled by local laws. Cyber laws in the context of India are still in the early developing stage. India has done many bilateral agreements related to Cyber-crime such as the agreement with Russia, a basic agreement with the US, and a framework cyber agreement with Israel to modernize it cyber space. But scope of these bilateral agreements is limited, ineffective and insufficient to deal cyber-crime. This research found that India should make multilateral treaty that blends its laws on common criminal policy and deal to reduce cyber-crimes at global level with international co- operation. This treaty will help in framing active regulations and strong analytical methods, which result in increasing internationally co-operation to control cyber-crime[45]. Budapest convention council of Europe on cybercrime is one of the multilateral international treaties that deal with international co-operation for fighting cyber-crimes globally. India should also join this Budapest Convention as US and Israel also whom India has bilateral agreement are already part of this cyber- crime  convention.

**Sarmah, A.; Sarmah, R. and Baruah, A. J.** (2017) found that with the development of new technologies are increasing; it also led to increase in crimes related to IoT[6]. Cyber-crime is actually a threat to people, so it is necessary to take steps to protect IoT from cyber-crimes for welfare of society, cultural and countries security aspect. Indian government has passed an IT act, 2000 to control cybercrimes issues. Cyber-crime is not bounded to any particular geographical area. It passes national boundaries on internet and creates legal and technical difficulties in investigation and accusing the crimes. Hence it is necessary that proper action should be taken to control cyber- crimes internationally. Proper cooperation and coordination among various countries are necessary to fight against cybercrimes. The main objective of this study is to create awareness among common people about Cyber-crime. It is important that users who become the victim of cyber-crime should come forward and report s file against cyber criminals so that strict actions can be taken against them and an example can be set for future cyber criminals.

**MS. Anisha**(2017) argue that main reasons of  cyber-crime  is  technology and dangerous infrastructure of Internet of Things (IoT). As the ratio of users on IoT is increasing, it leads to increase risks of different types of cyber-crimes. Crimes are unpredictable due to development of new technologies. Crimes based on technologies are increasing day by day and it is necessary that these are solved on the priority basis[7]. Cyber-crimes are not limited to computers only, electronic devices such as telecommunication tools, financial transaction machines etc. IoT is diversified in nature that makes it difficult to find out the problems of cyber security that results in unawareness on issues of security. It is suggested that free advertisements and workshops can be organized to generate awareness among users with the help of NGO'S and government. Cybercrimes issues and threats should be acknowledged from the grassroots level i.e. institutes, schools, colleges, computer centers etc. Indian government has taken many effective steps to control cybercrimes like Information technology act, 2000 has been passed. But a fixed cyber law is not effective in cyber-crime context as IoT has a vast and diversified nature and new type of crimes keeps on inventing. So, it is necessary that cyber law keep close eyes on cybercrime and updated its law and regulations accordingly.

**Husamuddin, M.; Qayyum, M.** (2017), explore that Internet of things are developing as an important technology. Information's transferred with RFID tags or sensors include sensitive data which should be secure from unauthorized access. There is no security in between 2 nodes of Internet of things communication and security of IoT should not be negotiated[8]. To get more secure communication it is necessary that IoT should have services like end to end environments, real time access control, protection of critical infrastructure and encryption. It is difficult to think or stay ahead of the cyber-criminal. It can be expected in future that smart equipment's will include privacy security which will help users to perform more tasks conveniently by using IOT. With improved privacy, information protected methods and ethical practices, IoT will gain the trust and faith of users in this connected world.

**Marion, N. E.** (2010) argues that the Council of Europe's (CoE) Agreement on Cybercrime is studied as representative mechanisms. This research shows that the Agreement includes the features of figurative policy like comforting the users that proper action taken to control cyber-crime, educate the users about cyber-crime, and act as a warning for those who commit cybercrimes. It leads to the serious consequences in IoT[9]. Crimes related to computers are not bound to any particular nation while it is an international matter. Countries should co-operate and make synchronized laws to have control on cybercrimes. The Coe treaty is such an important step towards cybercrime control globally. As Cybercrimes scope is very wide hence it is difficult to have proper control on it. In development of this treaty, representatives from different countries discuss and argue on the acts committed on internet and define the actions that can be taken to fight against cyber- crimes[33]. A reliable international method is used to control cybercrimes that include co-operation in law execution agencies and investigating offence. The finding shows that the effectiveness of CEO Treaty is questionable as resolutions made in this treaty are actually figurative like issues related to privacy, powers of investigators and it's kind of difficult to enforce co-operation among countries. There are many loopholes in this treaty that result in criminals to continue commit crimes. To have more effective and efficient treaty it is required that more and more countries sign this treaty and make national laws and keep on updating the laws as per new types of crimes.

**Moitra, S.** D (2005), explore in this research that there are various concerns that are very important for the development of policies related to cyber crimes[10]. Various concerns or issues are categorized in 5 questions that are 1.Criminals, 2.Crimes, 3.Occurrence of cyber-crime, 4.Effect on victims and 5.Response of society. It is also discussed in this research that why each concern is important in making policies. Europe Council enrolled its Agreement on Cybercrime in 2001, and various initiatives are launched by European Union to control cybercrime. Standardization and co-ordination that is necessary for common classification of cyber-crime is also discuss in this study. Hacker behavior, victim response, legal activity and criminal justice policies are some of the features that are to be studied thoroughly before making rules and regulations in cyber law. It is necessary to study cyber-crime further as new problems may occur in future and new types of cyber-crimes may appear, although policies still exist in various countries like in US, Agreement on Cybercrime (EU), Information technology act 2000 but up gradation is needed on timely basis. Finding shows that reliable information is to be collected and evaluated before developing new policies. Suggestions given in this research can help those countries as well which are still on development stage of making policies. This research has some limitations like it is limited to only some salient features and lots of other factors are still left.

**Oriwoh, E.; Sant, P. and Epiphaniou, G.** (2013) introduce some guideline principles for vendors, consumers, governments and law makers who use or work on IoT (Internet of Things)[11]. Usually new technologies and new applications of prevailing technologies show future prospects and appropriate uses where it can be used. Security concern importance in development of any technologies in IoT at early stage is acknowledged by various researchers. Laws are already in existence and new laws also provide guidance to the users of IoT and ensure that there should be no fraud or breach in usage of technologies; if it is detected then proper action should be taken to scold the offender[47]. Hence it is important that appropriate principles for guidance should be done to introduce the laws to the interesting users.

**Maung, T. M.; Thwin, M. M. S.** (2017), found that new challenges in forensic of cyber-crimes develop with new and latest operating systems[12]. On one side these new versions of Windows make the things easy for users and on the other possibilities of new crimes arise. Cyber-crime forensics investigations are not new arena and, but it is necessary that it keep updated its methods to catch cyber criminals. Computer forensics experts make cyber-crime investigations based on Quality, effectiveness, legal obligations and flexible. Objective of investigation must be customized, expertise, systematic and comprehensive enough that the process of investigation complete in less time and relevant information can be gathered and investigated accordingly[44]. Digital evidence in cyber-crime means the digital data that shows crime has committed and there is a relation between victim and crime or criminal and crime. IT Security is very difficult in this digital world as these exposed to various threats and malware's like viruses, spies, worms and Trojans affect IoT almost on daily basis[40]. This re- search shows various solutions that can be used systematically to get sound evidence forensically. These solutions help and support in collecting evident information in various forensic areas like cloud, static, and social network[25]. The main objective of this study is to find an appropriate solution for the country Myanmar.

## 3   TYPES OF CYBER ATTACKS ON IoT  DEVICES

### 3.1   Physical Attacks:

Physical attack s can happen when an IoT device is accessed physically. This kind of attack can be performed by the same company employee who has access to the IoT  device.

### 3.2   Encryption Attacks:

Encryption attack can be done when the IoT device is un-encrypted, attacker can sniff the data with the help of an intruder. Encryption attacks strike at the heart of your algorithmic system. Hackers analyze and deduce your encryption keys, to figure out how you create those algorithms. Once the encryption keys are unlocked, cyber-assailants can install their own algorithms and take control of your system.

### 3.3   DoS (Denial of Service) Attack:

This kind of attack may not steal the data from services like websites[24]. Attackers target service with a large number of botnets sending thousands of requests to the service and making them crash, which makes services unavailable.

### 3.4   Firmware Hijacking :

Firmware kind of attacks can be done when an IoT device is not updated up today. Attackers can hijack the device and download malicious software. Computers contain a lot of firmware, all of which is potentially vulnerable to hacking.

### 3.5   Botnet Attack:

A Botnet attack is which can be done when an IoT device is turned into remotely controlled bots, which can be used as a part of the botnet. Botnets can connect to the network and transfer the private and sensitive data. Two types of Botnet attacks a) The Mirai botnet b) The PBot malware

### 3.6   Man-in-the-Middle Attack:

A man-in-the-middle attack can be done when a hacker breather the communication between two systems. By spying on communication between two parties[17]. There are 7 types of Man-in-the-middle attack a) IP spoofing b) DNS spoofing c) HTTPS spoofing d) SSL hijacking e) Email hijacking f) Wi-Fi eavesdropping g) Stealing browser cookies

### 3.7   Ransomware Attack:

Ransomware is a type of attacks when a hacker encrypts the data and lock down the access. Then the hacker sells the decryption for his price. This kind of attack will disrupt daily business[51]. Types of Ransomware attack a) Scareware b) Screen lockers c) Encrypting ransomware.

### 3.8   Eavesdropping Attack:

Eavesdropping is a kind of attack when a hacker intercepts network traffic to get access to sensitive and private data via a weekend connection between an IoT device and a server[35]. Eavesdropping attack can prevent by using personal firewall, keeping antivirus software updated, and using a virtual private network.

### 3.9   Privilege Escalation Attack:

In this particular kind of attack the hacker looks for an IoT device bugs to gain access. In this attack, the use their newly gained privileges to deploy malware. hacker exploits a bug, design flow, or configuration error in an application or operating system to gain elevated access to the resources that should normally be unavailable to the third person.

### 3.10   Brute Force Password Attack:

In the attack, the hacker uses password hashing or password cracker software's to hit the server with the possible tries. Until the hacker gets the correct credentials. Hacker uses trail-and error method to guess Login credentials, encryption keys, or finding hidden web pages. Types of Brute force attacks: a) Simple brute force attacks. b) Dictionary attacks c) Hybrid brute force attacks d) Reverse brute force attacks e) Credential stuffing

## 4   SMART HOME AND IT'S SUBSYSTEMS

Smart homes are equipped with advanced automated internet connected devices which makes your life much easier like Multimedia kits, Automatic door and window operators, smart home appliances etc., but on the side of the coin, If one of the device of your smart home is vulnerable and unencrypted your data will be on Network like Google dorks, Shodan which the hackers first place to search[13]. The smart home appliances are mostly with three important entities: Physical Components, communication system and intelligent information processing. Which makes the most sophisticated smart home devices.

Adopting IoT Technology into our house will raise more security concerns and challengers, because the IoT based smart home devices are highly vulnerable to attack as compared to other devices[36]. If the Smart home device is attacked and compromised the potential to invade the user's privacy and steal the personal information and can be monitored as well[14].

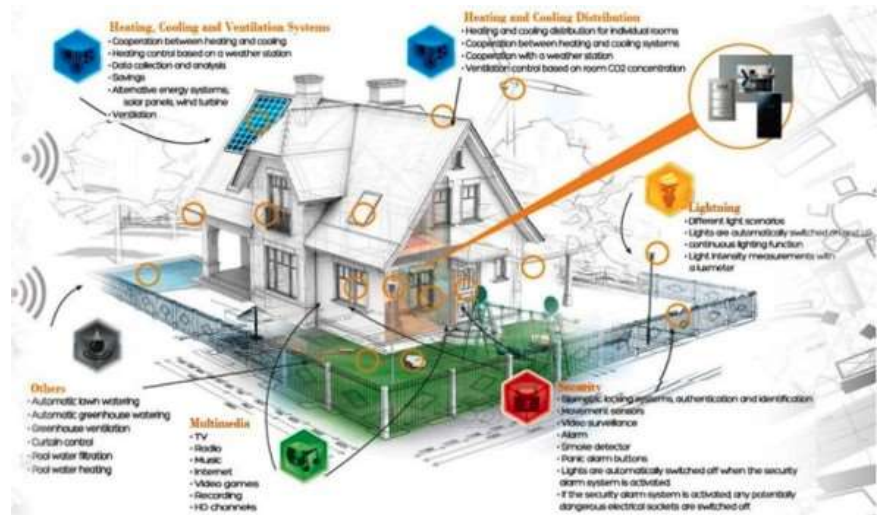

**Fig. 4.** Smart Home and Subsystems

# 5   CHALLENGES FOR THE CURRENT APPROACHES

## 5.1   Testing Drawbacks:

Due to insufficient testing and updating of current IoT devices i.e., Approx. 30 billion devices present connected to the internet, which makes the IoT devices vulnerable to cyber-attacks. Some IoT manufacturers offers firmware updates but unfortunately due to lack of automatic updates according to the zeroth day hacks, These Devices exposed to internet and prone by the hackers[43].

## 5.2   Default Passwords:

According to some the government reports that advised to the manufactures against selling the IoT devices with default credentials (admin / password)[46]. Weak passwords and default passwords of IoT devices are Identified as the most vulnerable devices with password hacking and brute-forcing.

## 5.3   IoT Ransomware:

The ransomware depends on encryption to lock out completely different users with different devices and platforms[34]. The ransomware could potentially protect the device from attack but eventually hacker can seal some personal data with the malware.

### 5.4   IoT AI and  Automation:

The amount of data gathered by the sensors and IoT Is becoming enormous day by day. AI tools and Automation are already been used in shifting massive information form network to network. However, using this AI tools for make autonomous decisions can affect Millions of functions across infrastructure such as like healthcare, transport etc., [29]  [48].

### 5.5   Botnet Attacks:

When a hacker creates a collection of malware infected botnet and sends thousands of requests per second and brings down the target is called as botnet attack[16]. A single IoT device infected with malicious code or malware sone not possess any real threat[19]. When the hacker uses DDoS attacks using thousands of IP cameras, home routers, Smart devices were infected and directed with bringing down the DSN provider platforms like Netflix, GitHub etc., [42].

## 6   RECOMMENDATIONS FOR CYBER HYGIENE OF IoT DEVICES

Installing updated firmware as soon as possible, where the manufacture released all the patches for the vulnerability's found in previous version of the device[52]. By updating the firmware one can be protected from the attacks by bugs from older versions of the software. It is recommended to change the pre-installed password or manufacture default passwords which makes your device default password attacks. Always reboot the device when you thin the device acting strange or when noticed suspicious activity[18]. It might help in getting rid of existing malware. keeping access to IoT device restricted by a local VPN can be protects from attacks.

## 7   CONCLUSION

IoT becomes the most important for every human being in this connected world. IoT (internet of things) make the world so small. Every human being is connected with each other by using internet. Lots of transactions are done over IoT and chances of crimes have increases. Crimes in cyber space are not bound to a particular location or country, its scope is very wide[30]. Risks related to Cyber-attacks can't be ignored so while adopting IoT on regular basis it is important to take adequate measures. Simple steps that can be done by users to get safe can be by installing software protection firewall, anti-viruses, anti- malware etc. Awareness should be spread regarding the types of cyber-crimes in IoT and how it can be avoided. Various international treaties are executed to control cyber-crimes. Current Multilateral and local legal devices and national laws include different concept and content, and their coverage of criminalizing, investigation powers and methods, digital evidence, risk and regulations, and control and co-operation internationally[36]. These treaties are different in different geographic scope like multilateral or regional and its applicability. These differences cause obstacles in identification, examination and taking action against cyber- criminals and taking preventive measures in cyber-crime. So, it is concluded that a common, cooperative, and flexible international treaties should be formed between the countries so that the same laws are followed everywhere on IoT and there is no confusion. Strict punishments should be imposed on cyber criminals so that potential criminals left the idea to do any crime. It should  ensure that laws formed must be followed properly and restrictions if any applied on internet access and content should not be neglected[31]. These laws should be according to the law and rights of humans[41]. Challenge is generated due to the scope and effect of cyber laws in different countries like the content on the internet is acceptable in one country while the same content is illegal in other countries than it is difficult to make common global laws.